\documentclass[conference]{IEEEtran}

\ifCLASSINFOpdf
\else
\fi
\usepackage{cite}
\usepackage{amsmath,amssymb,amsfonts}
\usepackage{graphicx}
\usepackage{textcomp}
\usepackage{xcolor}
\usepackage{lipsum}
\usepackage{multicol}
\usepackage{subcaption}
\usepackage{soul}
\usepackage{bm}
\usepackage{float}

\usepackage{algorithm,algpseudocode} 
\usepackage{amssymb, amsmath}
\usepackage{graphics}
\usepackage{url}
\usepackage{rotating}
\usepackage[compact]{titlesec}
\usepackage{pseudocode}  
\usepackage{algpascal}
\usepackage{algc} 
\usepackage{adjustbox}
\usepackage{epstopdf}
\usepackage{epsfig}
\usepackage{color}
\usepackage{multirow} 
\usepackage{url}  
\usepackage{caption}
\usepackage{tabularx}
\usepackage{textcomp}
\usepackage{amsmath}
\usepackage{soul}
\newcolumntype{Y}{>{\centering\arraybackslash}X}
\newcolumntype{K}[1]{>{\centering\arraybackslash}m{#1}}
\RequirePackage{enumitem}

\setlist[enumerate]{itemsep=0pt, topsep=0pt, itemindent=15pt, leftmargin=0pt,listparindent=\parindent}
\setlist[itemize]{itemsep=0pt, topsep=0pt, itemindent=15pt, leftmargin=0pt,listparindent=\parindent}

\def\BibTeX{{\rm B\kern-.05em{\sc i\kern-.025em b}\kern-.08em
    T\kern-.1667em\lower.7ex\hbox{E}\kern-.125emX}}
\newcommand{\ignore}[1]{}

\begin{document}

\title{Baseline  Drift Tolerant Signal Encoding for ECG Classification with Deep Learning}

\author{\IEEEauthorblockN{Robert O'Shea}
\IEEEauthorblockA{King's College London\\
Email: k1930297@kcl.ac.uk}
\and
\IEEEauthorblockN{Prabodh Katti}
\IEEEauthorblockA{King's College London\\
Email: prabodh.katti@kcl.ac.uk}
\and
\IEEEauthorblockN{Bipin Rajendran}
\IEEEauthorblockA{King's College London\\
Email: bipin.rajendran@kcl.ac.uk}}

\maketitle
\begin{abstract}

Common artefacts such as baseline drift, rescaling, and noise critically limit the performance of machine learning-based automated ECG analysis and interpretation. This study proposes Derived Peak (DP) encoding, a non-parametric method that generates signed spikes corresponding to zero crossings of the signal's first and second-order time derivatives. Notably, DP encoding is invariant to shift and scaling artefacts, and its implementation is further simplified by the absence of user-defined parameters. DP encoding was used to encode the 12-lead ECG data from the PTB-XL dataset ($n$=18,869 participants)  and was fed to  1D-ResNet-18 models trained  to identify myocardial infarction, conductive deficits and ST-segment abnormalities. Robustness to artefacts was assessed by corrupting ECG data with sinusoidal baseline drift, shift, rescaling and noise, before encoding.  The addition of these artefacts resulted in a significant drop in accuracy for seven other methods from prior art, while DP encoding maintained a baseline AUC of 0.88 under drift, shift and rescaling. DP achieved superior performance to unencoded inputs in the presence of shift (AUC under $1\,$mV shift: 0.91 vs 0.62), and rescaling artefacts (AUC 0.91 vs 0.79). Thus, DP encoding is a simple method by which robustness to common ECG artefacts may be improved for automated ECG analysis and interpretation.
\end{abstract}

\IEEEpeerreviewmaketitle

\section{Introduction}
Deep neural networks have demonstrated promising results for automated electrocardiograph (ECG) classification and monitoring \cite{Ebrahimi2020Sep}. However,  intensive energy requirements have been identified as a potential barrier to clinical implementation within the resource constraints of implanted and wearable devices \cite{Ebrahimi2020Sep}. Event-driven ECG processing limits communication to salient events such as waveform complexes, thereby reducing energy requirements \cite{Corradi, Yan2021Jan, banerjee2022snn, Bauer2019Nov, Chu2022Jul}. ``Spike''-based encoding optimises efficiency and latency further through ECG representation as sparse sequences of ``all-or-none'' spike signals. Spiking neural network models (SNNs) employ spike encoding throughout, facilitating accurate ECG classification on microjoule-range energy budgets\cite{Chu2022Jul}. Thus, SNNs present an opportunity to both improve battery-life and response latency in implanted cardiac defibrillators and pacemakers and wearable monitors \cite{Corradi, Yan2021Jan, banerjee2022snn, Bauer2019Nov}.
However, ECG signals are subject to significant distributional variability under normal operating conditions \cite{Lenis2017Mar, Chouhan}, presenting a major reliability obstacle for spike-based cardiovascular monitoring and diagnosis. Physiological factors such as respiration and limb movement cause baseline drift, a  low-frequency oscillatory artefact\cite{Perez-Riera2018Mar}. In contrast, diagnostic ECG features are subtle, as a $0.2\,$mV ST segment elevation may signify myocardial infarction \cite{Ibanez2018Jan}.  This problem is considered one of the most significant challenges for robust ECG classification \cite{Chouhan, Lenis2017Mar, Ebrahimi2020Sep, Das2018Mar}. Event-driven ECG encoding techniques typically convert analogue input to spikes according to fine-tuned threshold criteria \cite{Saeed2021Dec, Chu2022Jul, banerjee2022snn}. Consequently, these methods are particularly vulnerable to commonly encountered artefacts such as baseline drift \cite{Das2018Mar, Somani2021Aug}.   To improve the robustness of spike-based ECG encoding to these factors, we propose a parameter-free method based on zero-crossings of first and second-order time derivatives.

\subsection{Related Work}
 ``Rate" or ``count" encoding \cite{auge_hille_mueller_knoll_2021, Yan2021Jan}, encodes inputs over multiple channels, determining the proportion of active channels by input discretisation. ``Field response" encoding activates a single channel corresponding to the discretised input, improving efficiency by limiting instantaneous spike output to one-per-timestep in the channel dimension. ``Level crossing" encoding \cite{Saeed2021Dec, Chu2022Jul} generates spikes when discretised input values change, improving efficiency by omitting sequential spikes at similar levels \cite{Saeed2021Dec, Chu2022Jul}. ``BitString'' encoding or ``virtual neuron" \cite{Date2023Jul} encoding represents the discretised input as a signed $m$-bit integer over the channel dimension. BitString encoding encodes inputs as binary integer strings. As discretisation-based methods require pre-specified thresholds, parameterisation must balance adequate signal representation with energy efficiency and channel depth. All discretisation methods are theoretically vulnerable to baseline drift, which effectively shifts spiking thresholds relative to the ECG waveform. Scaling transformations affect the ratio of signal variance to discretisation interval widths, potentially leading to incorrect discretisation. Lastly, discretisation methods require multi-channel spiking, increasing model complexity.
 To address this problem, delta encoding\cite{Corradi, Bauer2019Nov}, also known as ``temporal contrast"  \cite{auge_hille_mueller_knoll_2021, Petro2019Apr}, yields a signed spike when the rate of change in the signal exceeds a predefined threshold. However, parameterisation may be challenging, and fixed thresholds may misrepresent data under gain changes. 

The ``Crossing Time-Encoding Machine" (CTEM) emits spikes when the input waveform intersects a regular reference waveform \cite{Gontier2014Jan}. The ``Crossing Time Encoding Machine with Feedback" (FTEM) introduces resetting of the reference waveform by the input to CTEM \cite{Gontier2014Jan}.  The integrate-and-fire time encoding machine (IFTEM) \cite{Lazar2004Jun, Naaman2022Apr} applies an integrate-and-fire neuron to the input waveform, emitting spikes when the membrane potential reaches a certain threshold. Due to the bipolar nature of ECG signal, for our baseline comparison studies,  we extend the IFTEM to have dual threshold such that both positive and negative spikes can be emitted. Some spiking neural network classifiers have employed recurrent architectures receiving ``direct input" of the unencoded  analogue current to the first layer of integrate-and-fire neurons \cite{Kovacs, DeMeloRibeiro2022Apr, Liang2022Jun}. This approach is functionally equivalent to real-time IFTEM, with empirical fitting of integrate and fire parameters.

\section{Derived Peak Encoding}
\begin{figure}
\centering
\includegraphics[width=0.49\textwidth]{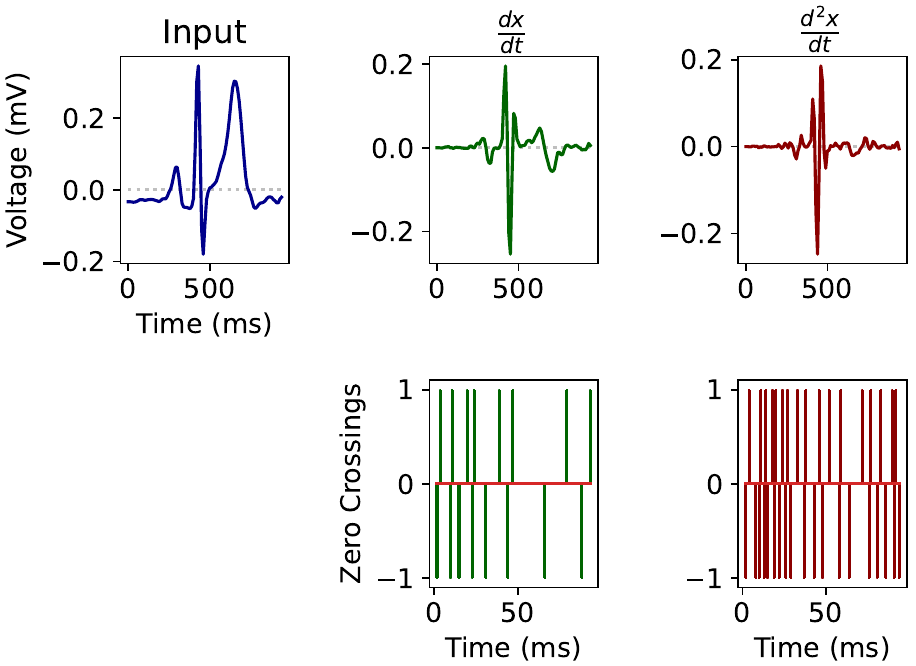}
    \caption{DP encoding scheme, for first and second-order derivatives, for a sample ECG.}
    \label{fig:dp}
\end{figure}

To address the vulnerabilities of threshold and discretisation-based encoding schemes, we propose ``Derived Peak" (DP) encoding for event-driven signalling. DP is physiologically inspired by  the peaks and nadirs which the P-wave, QRS complex and T-wave represent. Accordingly, DP generates signed spikes at zero-crossings of the signal's time derivatives such that:
\begin{equation}
S_{DP}(t)_k =\begin{cases}
1 & \frac{d^kx}{dt}(t)>0, \space \frac{d^kx}{dt}(t-1)\le0\\
-1 & \frac{d^kx}{dt}(t)<0, \space \frac{d^kx}{dt}(t-1)\ge0\\
0 & \text{otherwise}
\end{cases}
\label{eq5}
\end{equation}
Consequently, DP may be considered a variant of the CTEM class of encoding methods, with the distinctions that 1)  DP operates on the derivatives of the input to achieve location invariance, and 2) the zero constant is used as the crossing reference function, removing requirements for user-defined parameterisation. The zero crossings of the first derivative correspond temporally to peaks and nadirs of the input signal, while those of the second derivative correspond temporally to peaks and nadirs in the rate of change of the input signal (Fig. \ref{fig:dp}). Thus, it is observed that the information encoded by DP is directly related to the inflections after which the P-wave, QRS complex and T-wave are named\cite{Sattar2023Jun}. Furthermore, as $\operatorname{sgn}(\frac{d^kx}{dt})=\operatorname{sgn}(\frac{d^k(ax+b)}{dt})$ for all $k\ge 1$ and $a>0$, DP encoding is invariant to gain changes and shifts in the ECG waveform.

We propose two DP encoding schemes:
\begin{itemize}
    \item \emph{DP(1)}: DP encoding utilising the first order derivative of the signal ($k=1$ in equation \eqref{eq5}). The signal is presented to the network with 1 encoded channel per input channel.
    \item \emph{$DP(1,2)$}: DP encoding utilising both the first ($k=1)$ and the second order derivatives ($k=2$) of the signals. The signal is presented to the network with 2 encoded channels per input channel.
\end{itemize}
\section{Experiments and Data}
Electrocardiography data was extracted from the PTB-XL dataset via the PhysioNet repository \cite{Wagner2020May, Goldberger2000Jun}. 21,837 12-lead ECG records  sampled at 100 Hz were used in the study. 11,354 male and 10445 female patients were included, with median age 62 years (IQR: 50-72 years), median height 166 cm (IQR: 160-174 cm) and median weight 70 kg (IQR: 60-80 kg). 285 participants had pacemakers. Three (non-mutually exclusive) classes of cardiovascular conditions relating to ECG waveform morphology were annotated: ST-segment/T-wave changes (STTC), myocardial infarction (MI) and conduction deficits (CD). 5,235 samples (24\%) demonstrated STTC, 5,469 (25\%) demonstrated MI and 4,898 (23\%) demonstrated CD. Following the dataset authors' instructions, predefined folds 1-9 were used for model development, and fold 10 was reserved for testing.
\begin{figure}
\centering    
\includegraphics[width=0.48\textwidth]{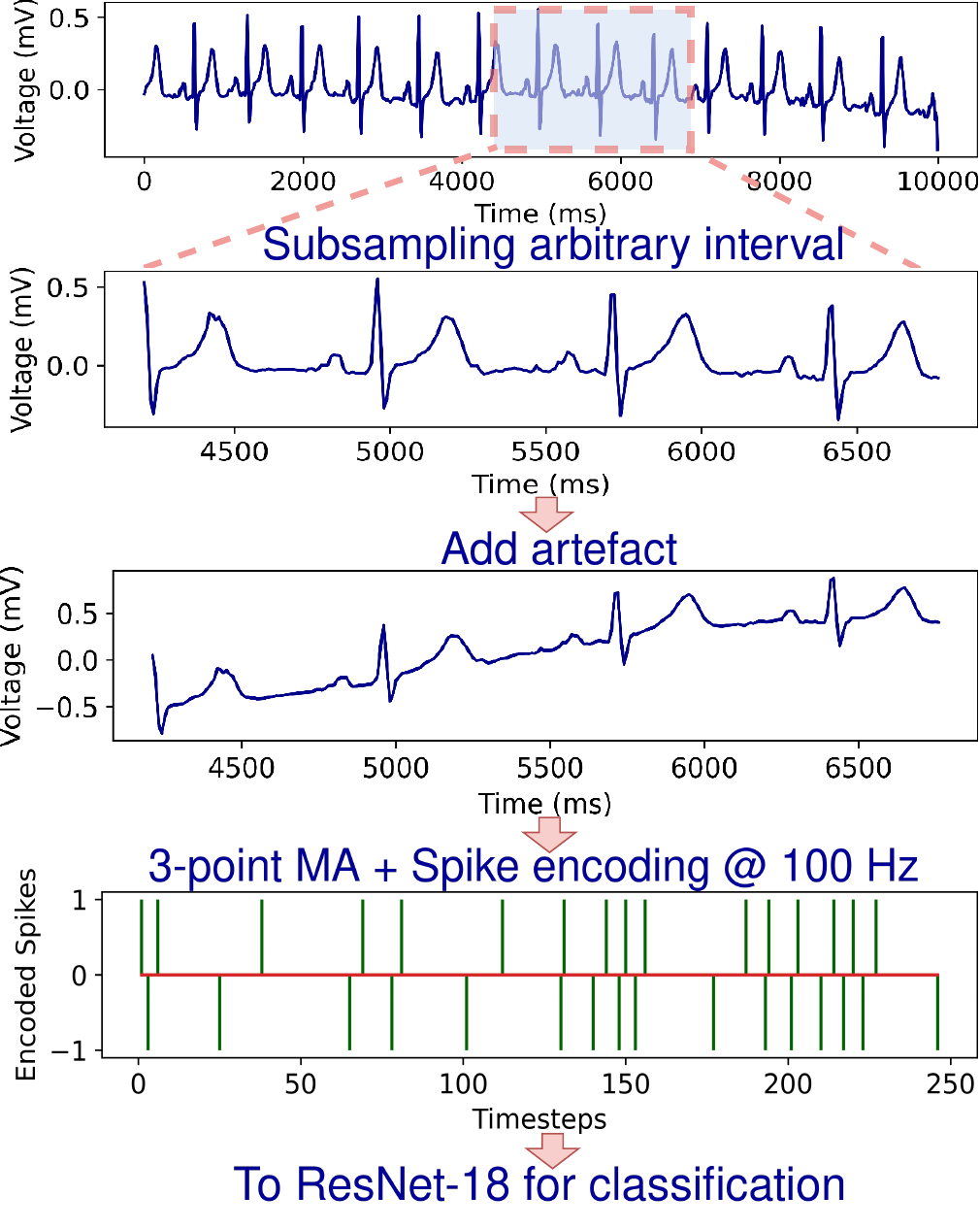}
    \caption{Experimental flow to test ResNet classification performance of spike-encoded ECG signals. To model the continuous low-latency ECG monitoring setting, subseries of 2,560ms duration were extracted from random timepoints. Signal pre-processing was limited to 3-point moving average (MA) for reduction of high-frequency noise.}
    \label{fig:experimental_flow}
  \end{figure}
The objective was to predict STTC, MI and CD from a 256-timestep, 12-lead ECG segment. A 256-timestep window was employed to assess the classification performance in the low latency setting, including approximately three beats in the temporal field. To recreate the most challenging and clinically useful scenario, models were provided with a random ECG segment, without waveform annotation or centering (Fig. \ref{fig:experimental_flow}).  Preprocessing was limited to a 3-point moving average, which served to reduce high-frequency noise. Event-driven encoding methods were applied to the ECG signal before it was passed to a 1D ResNet18 classifier for training/inference.  

Delta threshold parameterisation followed Corradi et al. \cite{Corradi}. Discretisation-based methods (Count, LC, FR, BitString) employed 31 bins of width $0.05\,$mV over the range $[-0.55, 1]$. CTEM was implemented with sine reference wave centered at $0\,$mV with $2\,$mV amplitude and 10-timestep wavelength ($10\,$Hz), for a target spike probability of $0.2$. FTEM was parameterised for a target spike probability of $0.2$ and  IFTEM was implemented with threshold, leakage and null bias targeting spiking probability of $0.05-0.1$.
\begin{figure}
\centering
\includegraphics[width=0.45\textwidth]{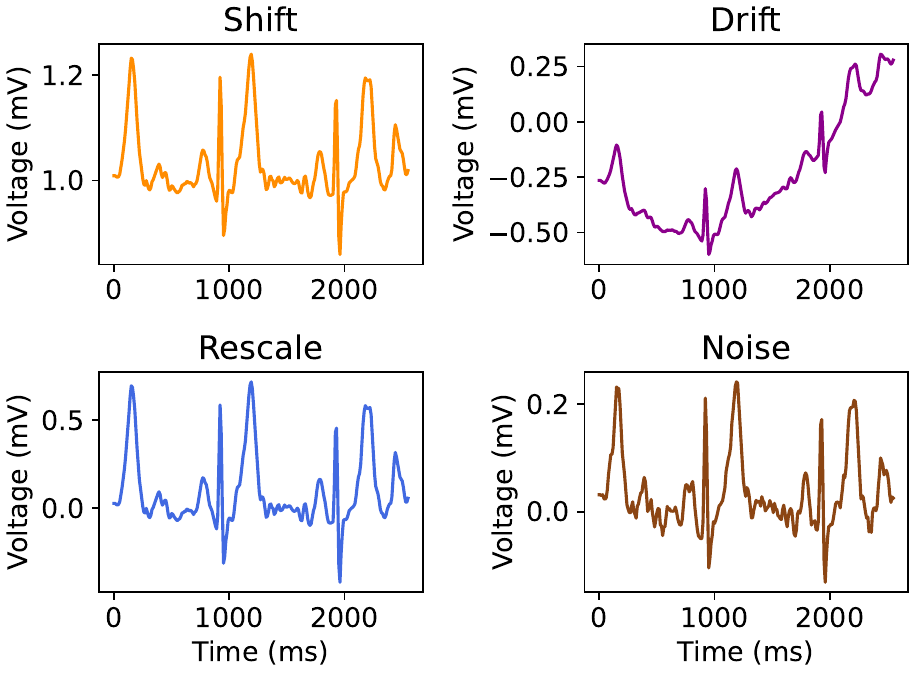}
    \caption{ECG sample corrupted with artefacts used in the study}
    \label{fig:noise}
\end{figure}
The training was performed with Tensorflow version 2.10 \cite{tensorflow2015-whitepaper}. Encoded inputs were passed to a 1D ResNet-18 architecture with a base feature depth of 64 convolutional filters. Global max-pooling was applied in the penultimate layer and passed to a 3-channel dense layer with sigmoid activation \cite{He}. 3-channel binary cross-entropy was minimised over batches of size 32, using the Adam optimizer with an initial learning rate of $10^{-3}$ and a decay factor of $10^{-7}$ per batch. Discrimination and calibration were evaluated on the test set using Area under Curve (AUC) and accuracy metrics. To assess robustness, further model testing was performed with four common artefacts (see Fig. \ref{fig:noise}):
\begin{enumerate}
\item Shift: $x(t)+a,\hspace{2mm} a=\pm1\,$mV
\item Rescale: $ax(t),\hspace{2mm} a\in\{\frac{1}{3}, 3\}$ 
\item Drift: $x(t) +\sin\left(\frac{(256+t+a)\pi}{512}\right),\hspace{2mm} a\in [0,256]$
\item Noise: $x(t) + \epsilon,\hspace{2mm} \epsilon \sim\mathcal{N}(0, 0.02)$
\end{enumerate}
Shift represents a change in the location of the ECG waveform baseline corresponding to an approximate QRS complex amplitude. Rescale represents a three-fold gain change. Drift is low frequency ($0.512\,$Hz) baseline drift oscillation with $1\,$mV amplitude. Noise represents spontaneous Gaussian interference with $10\%$ of signal standard deviation. 

\section{Results}
AUC and accuracy for each encoding method and artefact condition (averaged over diagnostic class) are provided in Table \ref{results_table}.  We highlight the following salient observations from our experiments:
\begin{itemize}
    \item In the absence of any artefact, the $DP(1,2)$ encoding schema yielded performance similar to all other methods, including unencoded inputs.
    \item In the presence of shift artefacts, models with DP(1) and  $DP(1,2)$ encoded inputs maintain stable calibrative (accuracy) and discriminative (AUC) performance. Encoding methods relying on predefined discretisation thresholds -- Rate, BitString, FR, LC -- as well as TEM-based methods failed under this perturbation. Delta encoding discriminative performance was robust to shift artefact.
    \item DP-encoded inputs maintained predictive performance in the presence of rescale artefact. All other methods, except for the one with unencoded inputs, suffered significant AUC and accuracy loss upon input rescaling. 
    \item DP-encoded inputs maintained predictive performance in the presence of drift artefact. The model receiving unencoded inputs was moderately robust to drift, as was delta encoding, but model performance with discretisation-based encoding and TEM-based methods reduced.
    \item Noise artefact induced mild deterioration in performance of models with DP(1) encoded inputs. However in the $DP(1,2)$-encoded case, model calibration decreased significantly.
\end{itemize}

\begin{table}
\centering
\begin{tabular}{K{0.7cm}K{1.1cm} K{1.0cm}K{0.6cm}K{0.9cm}K{0.7cm}K{0.7cm}}
\hline
\textbf{\emph{Metric}} & \textbf{\emph{Encoding}} & \textbf{\emph{Original}} & \textbf{\emph{Shift}} & \textbf{\emph{Rescale}} & \textbf{\emph{Drift}} & \textbf{\emph{Noise}} \\ \hline
\multirow{8}{*}{AUC} & None & \textbf{0.92}& 0.62& 0.79& 0.89& \textbf{0.92}\\ 
 & Rate & 0.91& 0.50& 0.77& 0.79& 0.91\\
 & BitString & 0.91& 0.50& 0.77& 0.79& 0.91\\
 & Delta & 0.89& 0.89& 0.75& 0.89& 0.89\\
 & FR & 0.90& 0.53& 0.74& 0.69& 0.90\\
 & LC & 0.91& 0.52& 0.81& 0.81& 0.91\\
 & IFTEM& 0.87& 0.62& 0.77& 0.83&0.86\\
 & CTEM& 0.87& 0.50& 0.8& 0.71&0.87\\
 & FTEM& 0.87& 0.49& 0.81& 0.80&0.87\\
 & \textbf{DP(1)}& 0.90& 0.90& 0.90& 0.90& 0.89\\
 & \textbf{DP(1,2)} & 0.91& \textbf{0.91}& \textbf{0.91}& \textbf{0.90}& 0.88\\ \hline
\multirow{8}{*}{Acc.} & None & \textbf{0.87}& 0.76& 0.80& 0.85& \textbf{0.87}\\ 
 & Rate & 0.87& 0.76& 0.78& 0.79& 0.87\\
 & BitString & 0.87& 0.55& 0.78& 0.75& 0.87\\
 & Delta & 0.86& 0.86& 0.76& 0.85& 0.86\\
 & FR & 0.86& 0.68& 0.73& 0.76& 0.86\\
 & LC & 0.87& 0.52& 0.78& 0.75& 0.87\\
 & IFTEM& 0.84& 0.60& 0.77& 0.81&0.84\\
 & CTEM& 0.84& 0.67& 0.80& 0.77&0.84\\
 & FTEM& 0.85& 0.75& 0.80& 0.80&0.85\\
 & \textbf{DP(1)} & 0.86& 0.86& 0.86& 0.85& 0.84\\
 & \textbf{DP(1,2)} & 0.86& \textbf{0.86}& \textbf{0.86}& \textbf{0.86}&0.81\\ \hline

\end{tabular}
\caption{Test AUC and accuracy of encoding schemes under each artefact condition}
\label{results_table}
\end{table}

Spike output rates produced by each method in the test dataset are provided in Table \ref{enc_table}. Rate and BitString encoding required multiple spikes per timestep. DP, TEM methods and LC generated relatively sparse outputs.  DP based methods achieved low spike counts throughout. Of the methods that outperform DP in efficiency, FR and LC have high output channel ratio, thereby increasing the model complexity.
\begin{table}
\centering

\begin{tabular}{K{1.2cm}K{1.4cm}K{1.4cm}K{1.4cm}}  
\hline
\textbf{\emph{Encoding}}& \textbf{\emph{Output Channel Ratio}}& \textbf{\emph{Total Channel Output}}&\textbf{\emph{Mean Channel Output}}\\
\hline
Delta& 1& 0.11&0.11\\
\hline
Rate& 31& 11.56&0.37\\
\hline
FR& 31& 1&0.03\\
\hline
LC& 31& 0.51&0.02\\
\hline
BitString& 7& 3.88&0.55\\
\hline
CTEM& 1& 0.21&0.21\\
\hline
FTEM& 1& 0.19&0.19\\ 
\hline
IFTEM& 1& 0.06&0.06\\ 
\hline
DP(1)& 1& 0.2&0.2\\
\hline
DP(1,2)& 2& 0.57&0.29\\
\hline
\end{tabular}
\caption{Spike output per 256-timestep ECG segment for various encoding methods, averaged over all test segments. The output channel ratio represents the total number of output channels per input channel. Total channel output is the total number of spikes across all channels, and mean channel output is the mean spikes per channel.}
\label{enc_table}
\end{table}

\section{Discussion}
Without artefacts, the classification accuracy obtained with a wide variety of spike-based encoding schemes is similar to that with analogue inputs. Test discrimination performance was comparable to that observed in a benchmark survey of deep learning model performance on the dataset $(AUC: 0.87-0.93)$, where the entire 1000-timestep series was provided to the model \cite{strodthoff2020deep}. This suggests that information loss that results due to event-driven encoding had a negligible effect on classification performance.

However, previously reported spike-encoding methods performed poorly in the presence of corrupting artefacts. Furthermore, models operating on the unencoded signal were also vulnerable to such corruption.  DP-encoded methods proved resilient to shift, rescale and drift artefacts, maintaining accuracy and AUC within a narrow 1-2\% range. An added advantage of DP encoding is its non-parametric nature, requiring minimal domain knowledge or extensive data distribution analysis to anticipate encoding behaviour and tuning.

However, high-frequency noisy interference presented a challenge for DP(1 ,2) based encoding, with DP(1) showing slightly better performance. High-frequency noise is known to complicate the computation of derivatives, and regularisation techniques have been proposed to address this problem \cite{vanBreugel2020}. Oscillations due to such noise in low amplitude regions in the signal may have caused artefactual zero crossings in the second-order time derivative.  Furthermore, diagnostic tasks which involve amplitude-based diagnostic criteria, such as hypertrophy \cite{Peguero2017Apr}, may be better served by amplitude-based encoding schemes.
Notably, DP  methods achieved high performance with relatively low spike rates, making them suitable candidates for deployment on energy-constrained diagnostic systems at the edge.

\section{Conclusion}

DP is a nonparametric, event-driven, and resource-friendly encoding technique which guarantees invariance to shifts and rescaling of the input signal with minimal pre-processing. DP encoded inputs yield similar classification performance to the unencoded input with improved robustness to three common artefacts. DP encoding has strengths and vulnerabilities complementary to those of currently available techniques, presenting an opportunity to improve cross-artefact robustness through hybrid implementation.

\section*{Acknowledgment}

 This research was supported in part by the EPSRC Open Fellowship EP/X011356/1.

\bibliography{references}
\end{document}